\documentclass[a4paper,12pt,english]{article}

\usepackage{amsfonts,bm,amssymb,euscript,array,babel}
\usepackage[nosort]{cite}
\usepackage{mathtools}
\usepackage{tikz-cd}
\usepackage{amsmath}
\usepackage{hhline}
\usepackage[amsmath]{empheq}
\usepackage{hyperref}
\usepackage{cancel}
\usepackage{fancybox}
\usepackage{arydshln}
\usepackage{mathrsfs}
\usepackage{pdfsync}
\usepackage{tikz-cd}

\newsavebox{\mysaveboxM}
\newsavebox{\mysaveboxT}

\makeatletter

\newcommand{\dd}{\mathrm{d}}

\newcommand{\e}{{\epsilon}}

\newcommand{\bbm}{\left(\begin{matrix}}
\newcommand{\ebm}{\end{matrix}\right)}
\newcommand{\beq}{\begin{eqnarray}}
\newcommand{\eeq}{\end{eqnarray}}

\newcommand{\cbral}{[\![}
\newcommand{\cbrar}{]\!]}
\newcommand{\Jac}{\text{Jac}}
\newcommand{\DFT}{double field theory}

\makeatother

\newtheorem{lemma}[equation]{Lemma}

\newtheorem{defn}[equation]{Definition}

\newtheorem{exa}[equation]{Example}

\newcommand{\sfrac}[2]{{\textstyle\frac{#1}{#2}}}

\newcommand{\be}{\begin{equation}}
\newcommand{\ee}{\end{equation}}

\newcommand{\beqa}{\begin{eqnarray}}
\newcommand{\eeqa}{\end{eqnarray}} 
\def\nn{\nonumber} \def \bea{\begin{eqnarray}} \def\eea{\end{eqnarray}}

\newcommand{\barr}{\begin{array}}
\newcommand{\earr}{\end{array}}
\numberwithin{equation}{section}

  \def\G{\Gamma}

    \def\r{\rho}
\def\s{\sigma}   

\def\cD{{\cal D}}

\def\cM{{\cal M}} \def\cN{{\cal N}}

 \def\one{\mbox{1 \kern-.59em {\rm l}}}

\def\bit{\begin{itemize}} \def\eit{\end{itemize}}

\def\({\left(} \def\){\right)}

\newenvironment{nalign}{
	\begin{equation}
		\begin{aligned}
		}{
		\end{aligned}
	\end{equation}
	\ignorespacesafterend
}

 \allowdisplaybreaks[3]

\begin{document}

\makeatother


\parindent=0cm

\renewcommand{\title}[1]{\vspace{10mm}\noindent{\Large{\bf

#1}}\vspace{8mm}} \newcommand{\authors}[1]{\noindent{\large

#1}\vspace{5mm}} \newcommand{\address}[1]{{\itshape #1\vspace{2mm}}}


\begin{titlepage}

\begin{flushright}
\small
RBI-ThPhys-2020-47
\end{flushright}

\begin{center}


\title{ {\Large 
{Double field theory algebroid and curved $L_\infty$-algebras}}}

\vskip 3mm

\authors{ 
\large
Clay James Grewcoe{\footnote{cgrewcoe@irb.hr}} and
Larisa Jonke{\footnote{larisa@irb.hr}}}

 \vskip 3mm
 
  \address{Division of Theoretical Physics, Rudjer Bo\v skovi\'c Institute \\ Bijeni\v cka 54, 10000 Zagreb, Croatia}

\begin{abstract}
\noindent
A double field theory algebroid (DFT algebroid) is a special case of the metric (or Vaisman) algebroid, shown to be relevant in understanding the symmetries of  double field theory.  In particular, a DFT algebroid  is a structure defined on a vector bundle over doubled spacetime equipped with the C-bracket of   \DFT.  In this paper we give the definition of a DFT algebroid as a curved $L_\infty$-algebra and show how implementation of the strong constraint of \DFT\ can be formulated as an $L_\infty$-algebra morphism. Our results provide a useful step towards coordinate invariant descriptions of \DFT\ and the construction of the corresponding sigma-model.
\end{abstract}

\end{center}

\vskip 2cm

\end{titlepage}

\setcounter{footnote}{0}

\section{Introduction and overview}\label{sec:intro}

Double field theory (DFT) is a field theory defined on a double, $2d$-dimensional  configuration space where the extra coordinates are introduced  in an effort to realise T-duality   of string theory as a symmetry of field theory  \cite{T,S1,S2,HZ1}. DFT  enjoys global $O(d,d)$  symmetry, and all objects in the theory belong to some representation of the $O(d,d)$ duality group.  In particular, coordinates and their duals form generalised coordinates $X^A=(x^a,\tilde x_a)$  in the fundamental representation of $O(d,d)$, and similarly derivatives $\partial_A=(\partial_a,\tilde\partial^a)$, while indices are raised and lowered by the $O(d,d)$ metric:
\be\eta_{AB}=\begin{pmatrix}
	0 & \delta^a_b\\
	\delta^b_a & 0
\end{pmatrix}~.\nn\ee
Throughout the paper $A,B,\ldots=1,\ldots,2d$ and  $a,b,\ldots=1,\ldots,d$ are  indices corresponding to double and standard spacetime respectively, while latin indices from the middle of the alphabet $I,J,\ldots=1,\ldots,2d$ and  $i,j,\ldots=1,\ldots,d$ are reserved for bundle indices. The field content of the theory, in its simplest form,  consists of the dilaton, the $d$-dimensional   metric $g_{ab}$ and the 2-form Kalb-Ramond field  $B_{ab}$, corresponding to the  universal gravitational massless bosonic sector of closed string theory. In principle, all fields in the theory depend on generalised coordinates.  The metric $g_{ab}$ and Kalb-Ramond field  $B_{ab}$ are packaged together into a symmetric {\it generalised metric} ${\cal H}_{AB}$
\be 
{\cal H}_{AB}=\begin{pmatrix} g_{ab}-B_{ac}\,g^{cd}\,B_{db} & -B_{ac}\,g^{cb} \nn\\ 
g^{ac}\,B_{cb} & g^{ab}
\end{pmatrix}~,
\ee
 or, in the frame formalism,  the $d$-dimensional bein $e^m{}_a$  and $B_{ab}$ are collected into  a generalised bein ${\cal E}^M{}_A$. The dilaton is combined with the determinant of the metric $g_{ab}$ into an   $O(d,d)$ scalar. However, from now on we shall drop  the dilaton field for simplicity, see \cite{Coimbra} for a geometric description of the dilaton field.

DFT also enjoys local, gauge symmetry combining standard diffeomorphisms and gauge transformations of   the Kalb-Ramond field, generated by   an  $O(d,d)$ vector $\epsilon^A={\cal E}_M{}^A\lambda^M =(\e^a,\tilde\e_a)$. However, the theory is gauge invariant only if the physical fields satisfy a set of constraints which can be written in a local basis as:
\be\label{sc}
\eta^{AB}\partial_A\partial_B(\,\cdots)=0~,\nn\ee
where dots in the bracket denote a product of different fields.   This \emph{strong constraint} (as opposed to the weak constraint of DFT that corresponds to the level-matching condition in string theory) also  enforces closure of the symmetry algebra of the theory \cite{HZ2}. Namely, the algebra of gauge transformation, $[\delta_{\e_1},\delta_{\e_2}]=\delta_{\e_{12}}$,  defined by the C-bracket, ${\e_{12}}=-\cbral\e_1,\e_2\cbrar$ 
\be\label{Cb}
\cbral\e_1,\e_2\cbrar^B=(\e_{1}^A\partial_A\e_{2}^B-\sfrac 12\eta_{AD}\e_{1}^A\partial^B\e_{2}^D)-(\e_1\leftrightarrow \e_2)~,\nn
\ee
closes only for fields and gauge parameters obeying the strong constraint. After imposing  the strong constraint, the C-bracket of DFT reduces to the Courant bracket, the properties of which are captured by  Courant algebroids \cite{C90,LWX,Pavol1}.

The question that naturally arises is if one can provide a geometric description of DFT symmetries based on the C-bracket,  before reducing the theory by imposing the strong constraint. In Ref.\cite{sd} the authors suggested that the relevant geometric structure is a pre-NQ manifold. This structure is defined on non-negatively (N) graded manifolds with a degree 1 vector field (Q) which does not square to zero, with the obstruction controlled by the strong constraint. The relevant pre-NQ manifold was obtained as a half-dimensional submanifold from the Vinogradov algebroid defined over a doubled space.

Similarly, motivated by the observation that in \DFT\ one doubles the configuration space, while in a Courant algebroid one extends the bundle, in  Ref.\cite{p1} the starting point was a ``large" Courant algebroid defined over a doubled target space.  Using a specific projection from the large Courant algebroid one arrives at a DFT algebroid,  the structural data of which are related to fluxes and the generalised bein of \DFT. After imposing the strong constraint, the DFT algebroid reduces  to a Courant algebroid over an undoubled target space. It was further shown that there exists a more general structure (dubbed a pre-DFT algebroid in \cite{p1}) that corresponds to the metric or Vaisman algebroid \cite{Vaisman}, of which the DFT algebroid is a special case. 

In this paper we shall analyse the structure of a DFT algebroid from a different perspective, giving its definition in terms of a \textit{curved} $L_\infty$-algebra \cite{csft,ls}. This naturally extends the results of \cite{sd}, and moreover, it implies that one should be able to formulate a DFT algebroid in terms of a Q structure. This becomes especially important when constructing the corresponding sigma-model, but this we leave for  future work.  
 Here we shall start by recalling the definition of a DFT algebroid and analysing its properties in more detail. The important observation is that the symmetric bilinear on the bundle  induces a symmetric pairing on the doubled configuration space, which allows for the appropriate geometric description of a DFT algebroid. This pairing is actually a para-Hermitan metric on the doubled configuration space, see Refs.\cite{f1,ds,f2,dd2,sz} for the description of \DFT\ in terms of para-Hermitian manifolds, and more generally in terms of Born geometry. Next, in  section \ref{sec:2} we introduce  curved $L_\infty$-algebras in our convention  and as a motivating example we review  the Courant algebroid $L_\infty$-algebra \cite{rw, p2}.  Thereafter, in subsections \ref{subsec:noL1} and \ref{subsec:withl1} we construct the  curved $L_\infty$-algebra for a DFT algebroid on two different  graded spaces underlying the $L_\infty$-algebra. 
 Section \ref{sec:morph} is dedicated to the understanding  of the strong constraint on the  DFT algebroid  as an $L_\infty$-morphism. We  begin by recalling the definition of  $L_\infty$-morphisms and then explicitly construct the map from a DFT algebroid to a Courant algebroid. Appendix \ref{app:flux} is intended as a  reminder to the reader of the explicit relation between DFT algebroid structures and   \DFT\ data. Finally,  appendices \ref{app:l1} and \ref{app:calcmorph} provide completeness for longer calculations of sections \ref{subsec:withl1} and \ref{subsec:morphnol1}.

\section{DFT algebroid}\label{subsec:dft}

A DFT algebroid was constructed in Ref.\cite{p1} using a specific projection from a Courant algebroid defined over doubled target space. In particular, it was shown that the projected bracket  is the C-bracket of \DFT, and the projected  bundle map $\r$ is related to the generalised bein.  The explicit relation between DFT algebroid structures and the flux formulation of \DFT \ using local basis  is reviewed in Appendix \ref{app:flux}.  Here we discuss global properties of a DFT algebroid, starting with the following  definition: 
 \begin{defn}\label{dftalg1}
{\rm ({\bf DFT algebroid \cite{p1}})} Let ${\cal M}$ be a $2d$-dimensional manifold. A  {DFT algebroid} 
 is a quadruple $(L,\cbral\, \cdot \,,\, \cdot \,\cbrar,\langle
\, \cdot \,,\, \cdot \,\rangle,\rho)$, where $L$ is a vector bundle of
rank $2d$ over ${\cal M}$ 
equipped with  a skew-symmetric bracket
$\cbral\, \cdot \,,\, \cdot \,\cbrar:\G(L)\otimes\G(L)\to\G(L)$,
  a
non-degenerate symmetric form $\langle
\, \cdot \,,\, \cdot \,\rangle:\G(L)\otimes\G(L)\to C^\infty({\cal M})$, and 
 a smooth bundle map $\rho:L\to T{\cal M}$,
 such that:
\begin{enumerate}
  	\item $\langle {\cal D}f,{\cal D}g\rangle=\sfrac 14\, \langle \dd f,\dd g\rangle~;$\label{axiom1}
  		\item \ $\cbral e_1,f\,e_2\cbrar=f\,\cbral e_1,e_2\cbrar+\big(\rho(e_1)f\big)\,e_2-\langle e_1,e_2\rangle\,{\cal D}f~;$ \label{axiom2}
  		\item \ $\langle \cbral e_3,e_1\cbrar+{\cal D}\langle e_3,e_1\rangle,e_2\rangle+\langle e_1,\cbral e_3,e_2\cbrar+{\cal D}\langle e_3,e_2\rangle\rangle=\rho(e_3)\langle e_1,e_2\rangle~;$\label{axiom3}
  		\end{enumerate}
for all $e_i \in \G(L)$ and $f,g\in C^{\infty}({\cal M})$, where ${\cal D}:C^\infty({\cal M})\to\G(L)$ is the derivative defined through $\langle {\cal D}f,e\rangle=\sfrac 12\, \rho(e)f$. 
\end{defn}
 From property 1 in Definition \ref{dftalg1} it follows that a pairing in the bundle $L$  induces a symmetric pairing on $T{\cal M}$:
\begin{align}\begin{split}\label{eq:etapairing}
		 &\eta:T{\cal M}\times T{\cal M}\to  C^\infty({\cal M})\\
		&\eta=\sfrac 12 \eta_{AB}\dd X^A\vee \dd X^B,\qquad A,B=1,\ldots,2d~,
	\end{split}
\end{align}
or, in other words, an $O(d,d)$ metric with  components:
\be\eta_{AB}=\begin{pmatrix}
	0 & \delta^a_b\\
	\delta^b_a & 0
\end{pmatrix},\qquad a,b=1,\ldots,d~.\label{eq:eta}\ee
Here we introduced the symmetric  tensor product   $u\vee v=u\otimes v+v\otimes u$, in analogy with the more standard wedge product.
Since $\eta_{AB}$  is invertible it also defines a symmetric 2-vector:
\begin{align}\begin{split}\label{eq:etainverse}
		 &\eta^{-1}:T^\ast{\cal M}\times T^\ast{\cal M}\to  C^\infty({\cal M})\\
		&\eta^{-1}= \sfrac 12\eta^{AB}\partial_A\vee \partial_B,\qquad A,B=1,\ldots,2d~.
	\end{split}
\end{align}
The action on functions is defined  via the natural contraction with 1-forms:
\bea
\eta^{-1}(df)=\iota_{\eta^{-1}}df=
\eta^{AB}\partial_Af\partial_B~.
\eea
Additionally, we define the action of the symmetric 2-vector $\eta^{-1}$   on a section $v=v^A\partial_A$ of $\Gamma(T{\cal M})$ using the Schouten-Nijenhuis bracket for symmetric vectors \cite{mdv} as follows:
\bea
\eta^{-1}(v):=[\eta^{-1},v]_{SN}=
\eta^{AB}\partial_Av^C\partial_C\vee\partial_B~.
\eea

We would now like to further explore the structural data of the DFT algebroid. First, one notices that  Im($\cal D$) is not in the kernel of map $\rho$:
\be\label{ker}
(\rho\circ {\cal D})f=\sfrac 12\eta^{-1}(df)~.
\ee
Next, one can show that map $\rho$ is not a homomorphism:
\begin{align} \label{prehomo}
\rho(\cbral e_1,e_2\cbrar)(f)-[\rho(e_1),\rho(e_2)](f)=&-{\sf SC}_{\r}(e_1,e_2)f~,\nn\\
{\sf SC}_{\r}(e_1,e_2)f{}:=&\,\sfrac 12\eta(\r(e_1),{\eta^{-1}(df)(\r(e_2))})-(e_1\leftrightarrow e_2)~,
\end{align}
where the second bracket on the lhs  in the first line is the standard Lie bracket of vector fields, and we introduced a shorthand notation for the rhs.
Furthermore, the  bracket does not satisfy the Jacobi identity, in fact we have:
\be \label{preJ}
\Jac(e_1,e_2,e_3):=\cbral\cbral e_1,e_2\cbrar,e_3\cbrar+\text{cyclic}={\cal D}
{\cal N}(e_1,e_2,e_3) + {\sf SC}_{\rm Jac}(e_1,e_2,e_3) ~,\ee
where ${\cal N}$ is 
\be \label{eq:N}
{\cal N}(e_1,e_2,e_3):=\sfrac 13\, \langle \cbral e_1,e_2\cbrar,e_3\rangle+\text{cyclic}~,
\ee
and  we introduced the shorthand notation:
\bea
&&{\sf SC}_{\rm Jac}(e_1,e_2,e_3):=\nn\\
&&\sfrac 12\r^{-1}\left\{(-\eta^{-1}(\r(e_2))(\eta(\r(e_1))+\eta^{-1}(\r(e_1))(\eta(\r(e_2))+[\r(e_1),\r(e_2)])\r(e_3)\right\}+\text{cyclic}~.\nn
\eea
Compared with definitions in \cite{p1}, the definition of ${\sf SC}_{\rm Jac}$ is extended, see App. \ref{app:flux} for more details. Here we defined the inverse of the anchor map $\rho^{-1}:T{\cal M}\to L$  as shown in the following commutative diagram:
\[ \begin{tikzcd}
L \arrow{r}{\hat{\eta}} &  L^*\\%
T\mathcal{M} \arrow[swap]{u}{\rho^{-1}}  \arrow{r}{\eta}& T^*\mathcal{M}  \arrow[swap]{u}{\rho^*} 
\end{tikzcd}\qquad :\qquad\hat{\eta}\circ\rho^{-1}=\rho^*\circ\eta~,
\]
where the map $\hat\eta:L\to L^*$ is induced by the DFT algebroid symmetric form. With a slight abuse of notation, we denote the bilinear form and the map it induces with the same letter, both for $\hat\eta$ and $\eta$. The existence of the inverse map $\r^{-1}$ is due to property 1 of Definition \ref{dftalg1} and the relation of the anchor to the generalised bein in DFT, as reviewed in appendix \ref{app:flux}.
 
 Finally,  in analogy to three further properties of  a Courant algebroid,  see Prop. 4.2 and Lemmas 5.1 and 5.2 of  \cite{rw}, three additional properties of a DFT algebroid will prove useful in the ensuing analysis.
 \begin{lemma}\label{eq:RoyProp} The following identities hold in a DFT algebroid:
\begin{enumerate}
	\item $2\langle e_1,\cD\langle e_2,\cD f\rangle\rangle-2\langle e_1,\cbral e_2,\cD f\cbrar\rangle  =\rho(\cD f)\langle e_1,e_2\rangle+{\sf SC}_{\r}(e_1,e_2)f$~, 
\item $\cN(e_1,e_2,\cD f)-\sfrac 14\r\cbral e_1,e_2\cbrar f=-\sfrac 14 {\sf SC}_{\r}(e_1,e_2)f$~, 
\item $\cN(\cbral e_1,e_2\cbrar,e_3,e_4)+\langle\cD\cN(e_1,e_2,e_3),e_4\rangle+\text{antisymm.}(1,2,3,4)=$ \\ $\hspace{3cm}=\sfrac 1 2 \langle{\sf SC}_{\rm Jac}(e_1,e_2,e_3),e_4\rangle+\text{antisymm.}(1,2,3,4)$~,\\
where $\text{antisymm.}(1,2,3,4)$ indicates all terms needed for the antisymmetrisation of $e_1$, $e_2$, $e_3$ and $e_4$. 
\end{enumerate}
\end{lemma}
The proof is easily obtained by direct calculation using the defining properties of a DFT algebroid.
Note that in the case of a Courant algebroid the rhs of all three properties above vanishes.

\section{$L_\infty$-algebra for the DFT algebroid}\label{sec:2}

Our aim in this section is to show that a DFT algebroid can  be understood as a \emph{curved} $L_\infty$-algebra.  We shall start, however, with a brief introduction to curved $L_\infty$-algebras.  

\subsection{On  (curved)  $L_\infty$-algebras}
$L_\infty$-algebras are generalizations of Lie algebras with infinitely-many higher brackets, related to each other by higher homotopy versions of the Jacobi identity \cite{csft,ls},  defined as follows.
\begin{defn}{\rm ({\bf $L_\infty$-algebra \cite{ls}})}
 A $L_\infty$-algebra  $(\mathsf{L},\mu_i)$ is a graded vector space $\sf L$ together with a collection of multilinear maps that are graded totally antisymmetric:
\be
\mu_i:{\sf L}^{\times i}\to \mathsf{L}, \nonumber
\ee
of degree $2-i$ where $i\in\mathbb{N}_0$ and  satisfy the homotopy Jacobi identities:
\be\sum_{j+k=n}\sum_\sigma\chi(\sigma;l_1,\ldots,l_n)(-1)^{k}\mu_{k+1}(\mu_j(l_{\sigma(1)},\ldots,l_{\sigma(j)}),l_{\sigma(j+1)},\ldots,l_{\sigma(n)})=0~;\label{eq:homotopyjac}\nn\ee
for all $l_i\in\mathsf{L}$,  $n\in\mathbb{N}_0$. Here $\chi(\sigma;l_1,\ldots,l_n)$ indicates the graded Koszul sign including the sign from the parity of the permutation of $\{1,\ldots,n\}$ that is ordered as: $\sigma(1)<\cdots<\sigma(j)$ and $\sigma(j+1)<\cdots<\sigma(n)$ (such permutations are also known as unshuffles).\label{linfty}
\end{defn}
The convention used is that totally graded antisymmetric means the following:
\[\mu_i(\ldots,l_r,l_s,\ldots)=(-1)^{|l_r||l_s|+1}\mu_i(\ldots,l_s,l_r,\ldots)~,\]
with $|l_r|$ the $\sf L$ degree of homogeneous element $l_r\in{\sf L}$.
 When $\mu_0\neq 0$   this algebra is called a curved $L_\infty$-algebra, while the name $L_\infty$-algebra usually refers to the case $\mu_0= 0$.  Importantly, when $\mu_0=0$, the map $\mu_1$  is a differential and the elements of the graded vector space $\mathsf{L}=\bigoplus_i\mathsf{L}_i$  form a cochain complex: 
\bea
\cdots   \overset{\mu_1}{\longrightarrow}   \mathsf{L}_{i} \overset{\mu_1}{\longrightarrow}   \mathsf{L}_{i+1} \overset{\mu_1}{\longrightarrow}     \cdots \nn
\eea
Here we would like to present two examples, relevant  in the context of this paper. 
\begin{exa}{\rm{\bf (Curved dgla \cite{clm})}
A curved differential graded Lie algebra is a triple $(\mathfrak{g},d,R)$ where $\mathfrak{g}$ is a graded Lie algebra, $d$ is a derivation with degree 1, and $R$ is a curvature element of degree 2 such that $dR=0$ and $d^2 x=[R,x]$ for all $x\in \mathfrak{g}$. In  the $L_\infty$ framework we identify   the map $\mu_0$ with the constant curvature $R$, $\mu_1$ with the derivation $d$ and $\mu_2$  with  the graded Lie bracket, satisfying the homotopy relations 
\begin{align*}
\mu_1\mu_0&=0~,\\
\mu_1(\mu_1(l))&=\mu_2(\mu_0,l)~.
\end{align*}}
\end{exa}
\begin{exa}{\rm{\bf (Courant algebroid \cite{rw,p2})} A Courant algebroid  is a quadruple ($E\to M, [\cdot\,,\cdot]_C,\langle\cdot\,,\cdot\rangle_C,a$) where $E$ is a   vector bundle  of rank $2d$ over a $d$-dimensional manifold $M$ with a skew-symmetric bracket defined on its sections, a symmetric bilinear and a bundle map $a$ to $TM$. All these structural data satisfy a certain number of properties or compatibility conditions, see e.g. \cite{C90,LWX,Pavol1}. Alternatively, a Courant algebroid  can be understood as an $L_\infty$-algebra where structural data are encoded in   maps on graded spaces, and the properties and compatibility conditions follow from the homotopy relations. We can choose the underlying graded vector space of the $L_\infty$-algebra as either the one in \cite{rw} (ignoring the space of constants $\mathsf{L}_{-2}$):
\[
\begin{array}{ccc}
{\sf L}_{-1} & \xrightarrow{D} & {\sf L}_0\\
f\in C^\infty(M)&  &e\in \Gamma(E)
\end{array}
\]
or the one proposed in  \cite{p2}:
\[
\begin{array}{ccccc}
{\sf L}_{-1} & \xrightarrow{D} & {\sf L}_0 &\xrightarrow{a}& {\sf L}_1\\
f\in C^\infty(M)&  &e\in \Gamma(E)&&h\in \mathfrak{X}(M)
\end{array}
\]
Here, $h=h^b\partial_b,\;b=1,\ldots,d$ and ${D}$ is the derivative defined through $\langle {D}f,e\rangle_{C}=\sfrac 12\, a(e)f$.
Maps that make this graded vector space an  $L_\infty$-algebra are given by the set (with $i\geqslant0$):
\begin{align}
\mu_1(f)&={D}f\nn\\
\mu_2(e_1,e_2)&=[ e_1,e_2]_C\nn\\
\mu_2(e,f)&=\langle e,{D}f\rangle_C\\
\mu_3(e_1,e_2,e_3)&={\cal N}_c(e_1,e_2,e_3)\nn\\\cdashline{1-2}
\mu_{i+1}(h_1,\ldots,h_i,e)&=h_1^{a_1}\cdots h_i^{a_i}\partial_{a_1}\cdots\partial_{a_i}a(e)^b\partial_b\nn
\end{align}
where ${\cal N}_c$ is defined in analogy to \eqref{eq:N} and the maps below the dashed line are the needed extension in the latter case, as shown in \cite{p2}. The extended structure is more natural in  case  we want to define a Courant algebroid   given the $L_\infty$-algebra, while Roytenberg and Weinstein have shown in \cite{rw} that starting from the Courant algebroid one can omit the ${\sf L}_1$ subspace from the graded space and uniquely define the $L_\infty$ structure. Furthermore, the properties of a Courant algebroid 
\begin{empheq}[box=\ovalbox]{align}\label{Cbox}
(a\circ D)f= 0\nn\\
a[e_1 ,e_2]_C-[a(e_1) ,a(e_2)]= 0\\
\Jac(e_1,e_2,e_3)-{D}
{\cal N}_c(e_1,e_2,e_3)=0\nn
\end{empheq}
come out in this form from the homotopy relations of the extended  $L_\infty$-algebra. Note that for the Courant algebroid  $\mu_0=0$, therefore the map $\mu_1$ is a differential and one can define the chain complex underlying the $L_\infty$-algebra structure \cite{rw }.  }
\end{exa}

 \subsection{Curved $L_\infty$-algebra for the DFT algebroid}\label{subsec:noL1}
 
The first proposal for  an $L_\infty$ structure relevant for DFT was given in Ref.\cite{sd} based on the  graded geometry  of a pre-NQ manifold.  In that case, the homotopy relations of the proposed $L_\infty$-algebra were satisfied only up to the strong constraint.  Here we wish to extend this result by constructing proper, albeit curved, $L_\infty$-algebra. 
We begin by defining the  relevant graded vector space:
\[
\begin{array}{ccccc}
\mathsf{L}_{-1}&\oplus  & \mathsf{L}_0 &\oplus &\mathsf{L}_2 \\
f\in C^\infty({\cal M}) & & e\in \Gamma(L) & &\mu_0
\end{array}\]
where $\mathsf{L}_2$ is a 1-dimensional vector space spanned by the constant element $\mu_0$. In general,  there is no chain complex underlying the graded vector space of curved $L_\infty$-algebras, thus we use $\oplus$ symbol connecting the spaces.
The maps that do not involve space $\mathsf{L}_2$ are taken in analogy with the Courant algebroid maps:
\begin{align}
\mu_1(f)&={\cal D} f~,\nn\\
\mu_2(e_1,e_2)&=\cbral e_1,e_2\cbrar\;,\;\mu_2(e,f)=\langle e,{\cal D}f\rangle~,\label{eq:flatmaps}\\
 \mu_3(e_1,e_2,e_3)&={\cal N}(e_1,e_2,e_3)~.\nn
\end{align}
 The maps involving $\mathsf{L}_2$ are going to be constructed from the homotopy relations. Before we begin with our construction, it is useful to see which homotopy relations will be non-trivial. To this end we prove the following
 \begin{lemma}
For every $l_1,l_2\in \mathsf{L}$ such that $\mu_i(l_1,l_2,\ldots)=0$, the homotopy Jacobi identities of Definition \ref{linfty} can be written in the following way:
\begin{align*}
\sum_{j+k=i}\sum_{\sigma'}\chi(\sigma';l_1,\ldots,l_i)(-1)^{k}\bigg(\mu_{k+1}(\mu_j(l_1,l_{\sigma'(2)},\ldots,l_{\sigma'(j)})l_2,l_{\sigma'(j+2)},\ldots,l_{\sigma'(i)})+{}\nn\\
+(-1)^{1+l_1l_2+(l_1-l_2)\sum_{m=2}^{j}l_{\sigma'(m)}}\mu_{k+1}(\mu_j(l_2,l_{\sigma'(2)},\ldots,l_{\sigma'(j)})l_1,l_{\sigma'(j+2)},\ldots,l_{\sigma'(i)})\bigg)&=0~.
\end{align*}\label{eq:mu0mu0}
\end{lemma}
The proof of the Lemma follows from the fact that since unshuffles are ordered and $l_1$ and $l_2$ must be in different products, all unshuffles will necessarily have  $l_1$ and $l_2$ for the first and $j+1$-st element or vice versa. This implies we can split the homotopy relation into two sums, those that have unshuffles that begin with $l_1$ and those that begin with $l_2$. Then it is simply a matter of connecting the graded Koszul signs of these two unshuffles. 

Additionally one can observe that in our case:
\be\label{eq:muimu0mu0} \mu_{i+1}(\mu_0,\mu_0,\ldots)=0,\qquad\forall i\in\mathbb{N}~, \ee
holds due to the graded antisymmetry of the maps and the fact that $|\mu_0|=2$. Therefore, all homotopy relations of two $\mu_0$ arguments must be trivial, which  reduces the number of identities to be calculated significantly.

We proceed by constructing the maps involving the  space  $\mathsf{L}_2$ from the homotopy relations. The homotopy identity  for $n=0$ is trivial in our case, so we move  on to $n=1$:
$$\mu_1\mu_1(l)= \mu_2(\mu_0,l)~.$$
This contains  one non-trivial identity:
\be\mu_2(\mu_0,e)=0~.\label{eq:mu0e}\ee
For $n=2$ the homotopy identity:
$$\mu_1(\mu_2(l_1,l_2))-\mu_2(\mu_1(l_1),l_2)-(-1)^{1+|l_1||l_2|}\mu_2(\mu_1(l_2),l_1)=-\mu_3(\mu_0,l_1,l_2)~,$$
contains three non-trivial cases: $(l_1,l_2)=\{(\mu_0,f),(e,f),(f_1,f_2)\}$. The first produces the condition:
\[
\mu_2(\cD f,\mu_0)=0~,
\]
which is automatically satisfied by \eqref{eq:mu0e}. The second and third are simply the definitions of higher brackets:
\begin{align}\label{a1}
\mu_3(\mu_0,e,f)&=\cbral e, \cD f\cbrar-\cD \langle e, \cD f \rangle\\
&=-\sfrac 12\rho^{-1}\eta^{-1}(df)(\r(e))~,\nn\\
\mu_3(\mu_0,f_1,f_2)&=2\langle\mathcal{D}f_1,\mathcal{D}f_2\rangle\label{a2}\\
&=\sfrac 1 2 \eta^{-1}(df_1,df_2)~.\nn
\end{align}

In the case of $n=3$,
\begin{align*}\mu_1(\mu_3(l_1,l_2,l_3))-\mu_2(\mu_2(l_1,l_2),l_3)+(-1)^{|l_2||l_3|}\mu_2(\mu_2(l_1,l_3),l_2)-&{}\\
-(-1)^{|l_1|(|l_2|+|l_3|)}\mu_2(\mu_2(l_2,l_3),l_1)+\mu_3(\mu_1(l_1),l_2,l_3)-&{}\\
-(-1)^{|l_1||l_2|}\mu_3(\mu_1(l_2),l_1,l_3)+(-1)^{|l_3|(|l_1|+|l_2|)}\mu_3(\mu_1(l_3),l_1,l_2)&=\mu_4(\mu_0,l_1,l_2,l_3)~,\end{align*}
there are four different  identities to be satisfied: $(l_1,l_2,l_3)=\{(\mu_0,e_1,e_2),(\mu_0,f_1,f_2),$ $(e_1,e_2,e_3),(e_1,e_2,f)\}$. The first case is a consistency condition satisfied due to \eqref{eq:mu0e} once one takes into account \eqref{eq:muimu0mu0}. For the next case, the identity is:
\[2\cD \langle \cD f_1,\cD f_2\rangle=-\mu_3(\mu_0,\cD f_1,f_2)-\mu_3(\mu_0,\cD f_2,f_1)~,\]
which is directly satisfied by use of \eqref{a1}. For choice $(e_1,e_2,e_3)$, the corresponding homotopy identity is a definition:
\[\cD\cN(e_1,e_2,e_3)-\Jac(e_1,e_2,e_3)=\mu_4(\mu_0,e_1,e_2,e_3)~,\]
or by use of \eqref{preJ}:
\begin{align} \mu_4(\mu_0,e_1,e_2,e_3) &=  -\mathsf{SC}_{\rm Jac}(e_1,e_2,e_3)~.\label{eq:mu0eee}
\end{align}
The last of the $n=3$ expressions defines $\mu_4(\mu_0,e_1,e_2,f)$:
\begin{align}\mu_4(\mu_0,e_1,e_2,f)&=-\langle\cbral e_1,e_2\cbrar,\cD f\rangle-\langle e_2,\cD\langle e_1,\cD f\rangle \rangle+\langle e_1,\cD\langle e_2,\cD f\rangle \rangle+\cN (\cD f, e_1, e_2)\nn\\
&=0~,\label{eq:mu0eef}\end{align}
where the second equality holds by \eqref{prehomo} and the first identity of Lemma \ref{eq:RoyProp}. 

Next is $n=4$ with three non-trivial conditions: $(l_1,l_2,l_3,l_4)=\{(\mu_0,e_1,e_2,f),(e_1,e_2,e_3,e_4),$ $(\mu_0,e,f_1,f_2)\}$. The first  case $(\mu_0,e_1,e_2,f)$ with condition:
\begin{align*}
\mu_4(\mu_1(f),\mu_0,e_1,e_2)+\mu_4(\mu_2(e_1,e_2),\mu_0,f)-\mu_3(\mu_2(e_1,f),\mu_0,e_2)+{}\\
+\mu_3(\mu_2(e_2,f),\mu_0,e_1)+\mu_2(\mu_3(\mu_0,e_1,f),e_2)-\mu_2(\mu_3(\mu_0,e_2,f),e_1)=0~,
\end{align*}
produces, after plugging in \eqref{eq:flatmaps} and \eqref{a1}:
\begin{align*}
\mathsf{SC}_{\rm Jac}(e_1,e_2,\cD f)+\cD \langle\cbral e_1,e_2\cbrar,\cD f\rangle+\cbral\cD f,\cbral e_1,e_2\cbrar\cbrar+{}\\+\cD \langle e_2,\cD\langle e_1,\cD f\rangle \rangle-\cD \langle e_1,\cD\langle e_2,\cD f\rangle \rangle-\cbral\cbral \cD f,e_1\cbrar,e_2\cbrar+\cbral\cbral \cD f,e_2\cbrar,e_1\cbrar=0~,
\end{align*}
which vanishes by use of \eqref{preJ} and \eqref{eq:mu0eef}. The second identity in $n=4$ is the definition:
\begin{align*}
\cN(\cbral e_1,e_2\cbrar,e_3,e_4)+\langle\cD\cN(e_1,e_2,e_3),e_4\rangle+\text{antisymm.}(1,2,3,4)=-\mu_5(\mu_0,e_1,e_2,e_3,e_4)~.
\end{align*}
Using the third identity of Lemma \ref{eq:RoyProp}, this can be rewritten as:
\begin{align}
\label{eq:mu0eeee}
\mu_5(\mu_0,e_1,e_2,e_3,e_4)&=-\sfrac 1 2 \langle{\sf SC}_{\rm Jac}(e_1,e_2,e_3),e_4\rangle+{\rm antisymm.}(1,2,3,4)~.
\end{align}
The last identity of $n=4$ is the compatibility:
\begin{align*}
\mu_3(\mu_2(e,f_1),\mu_0,f_2)+\mu_3(\mu_2(e,f_2),\mu_0,f_1)-\mu_2(\mu_3(\mu_0,e,f_1),f_2)-{}\\-\mu_2(\mu_3(\mu_0,e,f_2),f_1)-\mu_2(\mu_3(\mu_0,f_1,f_2),e)=0~.
\end{align*}
This can easily be shown to hold using  Lemma \ref{eq:RoyProp}. 

Moving on to $n=5$ with two non-trivial identities for: $(l_1,l_2,l_3,l_4,l_5)=\{(\mu_0,e_1,e_2,e_3,f),$ $(\mu_0,e_1,e_2,e_3,e_4)\}$. The first is:
\begin{align*}
0=\sfrac 1 2 \langle \mathsf{SC}_{\rm Jac}(e_1,e_2,e_3),\cD f\rangle+2\langle \cD\cN(e_1,e_2,e_3),\cD f\rangle+{}\\+\Big(\sfrac 1 2 \langle \mathsf{SC}_{\rm Jac}(e_1,e_2,\cD f ),e_3 \rangle+\cN(\cbral e_1,\cD f\cbrar, e_2, e_3)-\cN(\cD \langle e_1,\cD f\rangle, e_2,e_3)+\text{cyclic}(1,2,3)\Big)~,
\end{align*}
where by utilising properties  \eqref{prehomo}, \eqref{preJ} and Lemma \ref{eq:RoyProp},  one obtains:
\begin{align*}
\sfrac 1 6 \Big(\langle \mathsf{SC}_{\rm Jac}(e_1,e_2,\cD f),e_3\rangle+\mathsf{SC}_\rho(e_2,e_3)(\rho(e_1)f)+\text{cyclic}(1,2,3)\Big)+{}\\
+\sfrac 1 2 \langle \mathsf{SC}_{\rm Jac}(e_1,e_2,e_3),\cD f\rangle=0~.
\end{align*}
This relation can be shown to be identically satisfied by direct calculation. The second and last identity of $n=5$, after plugging in all the appropriate definitions, states:
\begin{align*}
-\mathsf{SC}_{\rm Jac}(\cbral e_1, e_2\cbrar,e_3,e_4)-\cbral e_4,\cD\cN (e_1,e_2,e_3)\cbrar+\cD\langle e_4,\cD\cN(e_1,e_2,e_3)\rangle+{}\\+\cbral\mathsf{SC}_{\rm Jac}(e_1,e_2,e_3),e_4\cbrar-\sfrac 1 2 \cD\langle\mathsf{SC}_{\rm Jac}(e_1,e_2,e_3),e_4\rangle+\text{antisymm.}(1,2,3,4)=0~.
\end{align*}
Properties \eqref{preJ} and the third identity of  Lemma \ref{eq:RoyProp} produce:
\[
\cbral\Jac(e_1,e_2,e_3),e_4\cbrar-\Jac(\cbral e_1,e_2\cbrar,e_3,e_4) +\text{antisymm.}(1,2,3,4)=0~,
\]
that is satisfied by direct calculation. 

Finally, $n=6$  has only one non-trivial relation $(l_1,l_2,l_3,l_4,l_5,l_6)=(\mu_0,e_1,e_2,e_3,e_4,e_5)$ where, by the definitions given above, one obtains:
\begin{align*}
\sfrac 1 2 \langle\mathsf{SC}_{\rm Jac}(\cbral e_1, e_2\cbrar,e_3,e_4),e_5\rangle-\cN(\mathsf{SC}_{\rm Jac}(e_1,e_2,e_3),e_4,e_5)-{}\\-\sfrac 1 2 \langle \cD\langle\mathsf{SC}_{\rm Jac}(e_1,e_2,e_3),e_4\rangle,e_5\rangle+\text{antisymm.}(1,2,3,4,5)=0~.
\end{align*}
A straightforward but lengthy and rather tedious direct calculation shows this holds. 

All higher homotopy identities vanish and we summarise our findings in the following boxed set of equations.
\begin{empheq}[box=\ovalbox]{align}\label{box}
\mu_1(f)&={\cal D} f\nn\\
\mu_2(e_1,e_2)&=\cbral e_1,e_2\cbrar\nn\\
\mu_2(e,f)&=\langle e,{\cal D}f\rangle\nn\\
\mu_3(e_1,e_2,e_3)&={\cal N}(e_1,e_2,e_3)\nn\\
\mu_3(\mu_0,e,f)&=\cbral e, \cD f\cbrar-\cD \langle e, \cD f \rangle\\
\mu_3(\mu_0,f_1,f_2)&=2\langle\mathcal{D}f_1,\mathcal{D}f_2\rangle\nn\\
\mu_4(\mu_0,e_1,e_2,e_3) &= \cD\cN(e_1,e_2,e_3) -\Jac(e_1,e_2,e_3)\nn\\
\mu_5(\mu_0,e_1,e_2,e_3,e_4)&=\sfrac 1 2 \langle\cD\cN(e_1,e_2,e_3),e_4\rangle-\sfrac 1 2 \langle\Jac(e_1,e_2,e_3),e_4\rangle+{}\nn\\
&\phantom{\,=\,}+\text{antisymm.}(1,2,3,4)\nn
\end{empheq}
All non-zero maps that include the constant element $\mu_0$ of the space $\mathsf{L}_2$  are controlled by the pairing on $T\cM$ \eqref{eq:etapairing} and its inverse \eqref{eq:etainverse}, as can be seen from \eqref{ker} and   Lemma \ref{eq:RoyProp}. Here we choose to represent the space $\mathsf{L}_2$ as the  space spanned by the  constant symmetric bivector $\eta^{-1}$.

\subsection{Extending the curved $L_\infty$-algebra for the DFT algebroid}\label{subsec:withl1}

A better understanding of the  DFT algebroid that arises from the $L_\infty$ structure can be obtained if we extend the underlying vector space by adding $\mathsf{L}_1$,  containing  sections of $T\cal M$. In that way, the anchor map is included in the $L_\infty$ maps, the choice of representation of $\mathsf{L}_2$ as the space spanned by the constant symmetric bivector $\eta^{-1}$ is natural, and the homotopy relations reproduce the defining properties of  a DFT algebroid. Therefore, we shall start with the following graded vector space:
\[
\begin{array}{ccccccc}
\mathsf{L}_{-1}&\oplus  & \mathsf{L}_0&\oplus   & \mathsf{L}_1 &\oplus  & \mathsf{L}_2  \\
f\in C^\infty({\cal M}) & & e\in \Gamma(L) & &  h\in \mathfrak{X}(\mathcal{M})& &{\mu_0}
\end{array}\]
the boxed maps \eqref{box} and 
\begin{align}\label{eq:he}
\mu_{i+1}(h_1,\ldots,h_{i},e)=h_1^{A_1}\cdots h^{A_i}_i\partial_{A_1}\cdots\partial_{A_i}\rho(e)^B\partial_B~,\qquad i\geqslant 0~,
\end{align}
 a choice based on the analogous  relation for   Courant algebroids. Additional maps are  constructed from the  homotopy identities as follows.

As in the previous subsection we begin our construction with the $n=1$ homotopy identity since the $n=0$ case is trivial. This case has two non-trivial possibilities $l=f$ and $l=e$. The first produces:
\be\mu_2(\mu_0,f)=\mu_1(\mu_1(f))=\rho\circ\mathcal{D}f=\sfrac 1 2 \eta^{AB}\partial_Bf\partial_A~,\ee
whereas the second:
\[\mu_1\rho(e)=\mu_2(\mu_0,e)\in\mathsf{L}_2~,\]
must be trivial since $\mathsf{L}_2$ is by construction spanned by the constant element $\mu_0$ and cannot, therefore, non-trivially depend on an arbitrary section $e$ of $L$. Thus the following must hold:
\be\mu_1(h)=0\quad\text{and}\quad\mu_2(\mu_0,e)=0~.\ee
It is interesting to note that, since there is an ${\sf L}_1$ space in this extension, one can explicitly see the curving of our ``differential'' $\mu_1$ on functions. The same reasoning implies that all homotopy identities in the space $\mathsf{L}_2$ must be trivially satisfied:
\begin{align}
	\begin{split}
		\mu_i(h_1,\ldots,h_i)&=0~,\\
		\mu_{i+2}(h_1,\ldots,h_i,\mu_0,e)&=0~.
	\end{split}
\end{align}
Using Lemma \ref{eq:mu0mu0} one can show that in general we can have at most 15 non-trivial identities for each $n$. 

Moving on to $n=2$, we find  4 non-trivial identities, however, only three of these are different from the $\mathsf{L}_1=\varnothing$ case above: $(l_1,l_2)=\{(e,f),(e_1,e_2),(h,f)\}$. These give, respectively:
\begin{align*}
\mu_2(\rho(e),f)&=0~,\\
\rho\cbral e_1,e_2\cbrar-[\rho(e_1),\rho(e_2)]&=-\mu_3(\mu_0,e_1,e_2)~,\\
\rho\mu_2(h,f)+\sfrac 1 2 \eta^{BC}h^A\partial_A\partial_Cf\partial_B&=-\mu_3(\mu_0,h,f)~,
\end{align*}
that result in:
\begin{align}\begin{split}
		\mu_2(h,f)&=0~,\\
		\mu_3(\mu_0,e_1,e_2)&=\mathsf{SC}_\rho(e_1,e_2)~,\label{eq:mu3}\\
		\mu_3(\mu_0,h,f)&=-\sfrac 1 2 \eta^{BC}h^A\partial_A\partial_Cf\partial_B~.
	\end{split}
\end{align}

Continuing to the $n=3$ case, one has 8 non-trivial identities, of these only 5 are new in comparison to the previous subsection. They are: $(l_1,l_2,l_3)=\{(h,f_1,f_2),(h,e,f),$ $(h_1,h_2,f),(e_1,e_2,h),(\mu_0,e,f)\}$ with the corresponding homotopy expressions:
\begin{align*}
\mu_4(\mu_0,h,f_1,f_2)&=0~,\\
\mu_4(\mu_0,h,e,f)&=0~,\\
\mu_3(\mathcal{D}f,h_1,h_2)&=\mu_4(\mu_0,h_1,h_2,f)~,\\
\mu_2(h,\cbral e_1,e_2\cbrar)+\mu_2(\mu_2(e_1,h),e_2)+\mu_3(\rho(e_1),e_2,h)-e_1\leftrightarrow e_2&=\mu_4(\mu_0,e_1,e_2,h)~,\\
\rho\cbral e,\cD f\cbrar-i_{\eta^{-1}}d\langle e,\cD f\rangle+\sfrac 1 2\mu_2( i_{df}\eta^{-1},e)-{}\\
-\mu_2(\langle e,\cD f\rangle,\mu_0)-\mu_3(\rho(e),\mu_0,f)+\mu_3(\cD f,\mu_0,e)&=0~.
\end{align*}
The first four are definitions of higher maps:
\begin{align}\begin{split}
		\mu_4(\mu_0,h_1,h_2,f)&=\sfrac 1 2 \eta^{BC}h^{A_1}_1h^{A_2}_2\partial_{A_1}\partial_{A_2}\partial_Cf\partial_B~,\\
		\mu_4(\mu_0,e_1,e_2,h)&=-h^A\partial_A\mathsf{SC}_\rho(e_1,e_2)^B\partial_B~,\label{eq:mu4}
	\end{split}
\end{align}
whereas the last is a condition satisfied by use of \eqref{prehomo}, and the maps defined thus far. Definitions \eqref{eq:mu3} and \eqref{eq:mu4} suggest, in the spirit of \eqref{eq:he}, the following Ansatz for the non-vanishing maps:
\begin{align}\begin{split}\label{eq:mu0fmu0ee}
\mu_{i+2}(h_1,\ldots,h_i,\mu_0,f)&=\sfrac 1 2 \eta^{BC}h^{A_1}_1\cdots h^{A_i}_i\partial_{A_1}\cdots\partial_{A_i}\partial_Cf\partial_B~,\\
\mu_{i+3}(h_1,\ldots,h_i,\mu_0,e_1,e_2)&=h^{A_1}_1\cdots h^{A_i}_i\partial_{A_1}\cdots\partial_{A_i}\mathsf{SC}_\rho(e_1,e_2)^B\partial_B~.\end{split}
\end{align}
Using this Ansatz one can show that all higher identities, which are infinite in number, are satisfied, see appendix \ref{app:l1}. We collect the maps for the extended $L_\infty$-algebra corresponding to a DFT algebroid in the following list (where $i\geqslant0$).
\begin{align}\begin{split}
		\mu_1(f)&={\cal D} f\\
		\mu_2(e_1,e_2)&=\cbral e_1,e_2\cbrar\\
		\mu_2(e,f)&=\langle e,{\cal D}f\rangle\\
		\mu_3(e_1,e_2,e_3)&={\cal N}(e_1,e_2,e_3)\\
		\mu_3(\mu_0,e,f)&=\cbral e, \cD f\cbrar-\cD \langle e, \cD f \rangle\\
		\mu_3(\mu_0,f_1,f_2)&=2\langle\mathcal{D}f_1,\mathcal{D}f_2\rangle\\
		\mu_4(\mu_0,e_1,e_2,e_3) &= \cD\cN(e_1,e_2,e_3) -\Jac(e_1,e_2,e_3)\\
		\mu_5(\mu_0,e_1,e_2,e_3,e_4)&=\sfrac 1 2 \langle\cD\cN(e_1,e_2,e_3),e_4\rangle-\sfrac 1 2 \langle\Jac(e_1,e_2,e_3),e_4\rangle+{}\\
		&\phantom{\,=\,}+\text{antisymm.}(1,2,3,4)\\
		\mu_{i+1}(h_1,\ldots,h_i,e)&=h_1^{A_1}\cdots h_i^{A_i}\partial_{A_1}\cdots\partial_{A_i}\r (e)^B\partial_B\\
		\mu_{i+2}(h_1,\ldots,h_{i},\mu_0,f)&=\sfrac 1 2 \eta^{BC}h_1^{A_1}\cdots h_i^{A_i}\partial_{A_1}\cdots\partial_{A_i}\partial_Cf\partial_B\\
		~~\mu_{i+3}(h_1,\ldots,h_i,\mu_0,e_1,e_2)&=h_1^{A_1}\cdots h_i^{A_i}\partial_{A_1}\cdots\partial_{A_i}{\sf SC}_\rho(e_1,e_2)^B\partial_B
	\end{split}
\end{align}
The homotopy relations reproduce the defining properties of a DFT algebroid   as discussed in Sect. \ref{subsec:dft}:
\begin{empheq}[box=\ovalbox]{align}
(\r\circ {\cal D})f&= \sfrac 12 \eta^{-1}(df)\nn\\
\r\cbral e_1 ,e_2\cbrar_C-[\r(e_1) ,\r(e_2)]&= -{\sf SC}_\rho(e_1,e_2)\\
\Jac(e_1,e_2,e_3)-{{\cal D}}
{\cal N}(e_1,e_2,e_3)&={\sf SC}_{\rm Jac}(e_1,e_2,e_3)\nn
\end{empheq}
and their higher derivatives.

\section{$L_\infty$-morphism as the strong constraint}\label{sec:morph}

In order to complete the description of a DFT algebroid in terms of an $L_\infty$-algebra, we would also like to include  the strong constraint in this framework. Since we know that on the solution of the strong constraint the  C-bracket of \DFT\  reduces to the Courant bracket, we are looking for a relation between the  $L_\infty$-algebra for a DFT algebroid and the one for a Courant algebroid. The natural relation between $L_\infty$-algebras is an $L_\infty$-algebra morphism or $L_\infty$-morphism for short. In the following, we explicitly construct an $L_\infty$-morphism from DFT to a Courant algebroid implementing the strong constraint. 
 
\subsection{On (curved)  $L_\infty$-morphisms}
 Before we start with the construction of mappings, we first recall the definition of an $L_\infty$-morphism.
 \begin{defn}\label{linfmorph}{\rm ({\bf $L_\infty$-morphism  \cite{stasheff}})} A collection of multilinear, totally graded antisymmetric homogeneous maps $\phi_i:{\sf L}^{\times i}\to{\sf L}'$ of degree $1-i$, $i\in\mathbb{N}_0$, is an $L_\infty$-morphism between two $L_\infty$-algebras $({\sf L},\mu)$ and $({\sf L}',\mu')$ if they satisfy:
\begin{align*}
&\sum_{j+k=n}\sum_{\sigma\in \mathrm{Sh}(j;n)}(-1)^{k}\chi(\sigma;l_1,\ldots,l_n)\phi_{k+1}(\mu_j(l_{\s(1)},\ldots,l_{\s(j)}),l_{\s(j+1)},\ldots,l_{\s(n)})=\nn\\
&= \sum_{k_1+\cdots+k_j=n}\sfrac {1}{j!} \sum_{\sigma\in \mathrm{Sh}(k_1,\ldots,k_{j-1};n)}\chi(\sigma;l_1,\ldots,l_n)\zeta(\sigma;l_1,\ldots,l_n)\times \\
&\phantom{\,=\,}{}\times \mu'_j(\phi_{k_1}(l_{\s(1)},\ldots,l_{\s(k_1)}),\ldots,\phi_{k_j}(l_{\s(k_1+\hdots+k_{j-1}+1)},\ldots, l_{\s(n)})) ~,
\nn
\end{align*}
where $\chi(\sigma;l_1,\ldots,l_n)$  is the graded Koszul sign and $\zeta(\sigma;l_1,\ldots,l_n)$ the sign coming from the mixing of the degrees of the various maps $\phi_{k_i}$, given for a $(k_1,\ldots,k_{j-1};n)$-shuffle $\s$ by:
\bea
\zeta(\sigma;l_1,\ldots,l_n)=(-1)^{\sum_{1\leqslant r<s\leqslant j}k_rk_s+\sum_{r=1}^{j-1}k_r(j-r)+\sum_{r=2}^j(1-k_r)
	\sum_{k=1}^{k_1+\hdots+k_{r-1}}|l_{\s(k)}|} ~.\nn\eea
\end{defn}
Similarly to the expression for the homotopy relations, the condition of Def. \ref{linfmorph} is actually a  possibly infinite series of relations, one for each $n\in\mathbb{N}_0$. Here we explicitly state the first three: 
\begin{itemize}
	\item  {$n=0$}
	\begin{align*}
	\phi_1(\mu_0)=\mu_0'+\mu_1'(\phi_0)+\sfrac{1}{2!}\mu_2'(\phi_0,\phi_0)+\cdots
	\end{align*}
	\item  {$n=1$}
	\begin{align*}
	\phi_1(\mu_1(l))-\phi_2(\mu_0,l)=\mu_1'(\phi_1(l))+\mu_2'(\phi_0,\phi_1(l))+\sfrac{1}{2!}\mu_3'(\phi_0,\phi_0,\phi_1(l))+\cdots
	\end{align*}
	\item  {$n=2$}
	\begin{align*}
	&\phi_3(\mu_0,l_1,l_2)-\phi_2(\mu_1(l_1),l_2)+(-1)^{l_1l_2}\phi_2(\mu_1(l_2),l_1)+\phi_1(\mu_2(l_1,l_2))={}\\
	&=\mu_1'(\phi_2(l_1,l_2))+\mu_2'(\phi_0,\phi_2(l_1,l_2))+\sfrac{1}{2!}\mu_3'(\phi_0,\phi_0,\phi_2(l_1,l_2))+\cdots+{}\\
	&\phantom{\,=\,}+\mu_2'(\phi_1(l_1),\phi_1(l_2))+\mu_3'(\phi_0,\phi_1(l_1),\phi_1(l_2))+\sfrac{1}{2!}\mu_4'(\phi_0,\phi_0,\phi_1(l_1),\phi_1(l_2))+\cdots
	\end{align*}
\end{itemize}
In the case of  non-vanishing $\phi_0$ this is called a  {curved} $L_\infty$-morphism.

\subsection{From a DFT algebroid to a Courant algebroid}\label{subsec:morphnol1}

To set the stage, we begin with $\sf L$, a DFT algebroid over a doubled space $\cal M$, and $\sf L'$, a Courant algebroid over $M$, where $M$ is a subspace of $\cal M$ and $\dim M=\dim {\cal M}/2$. Then introduce a mapping $\phi:{\sf L}\to{\sf L'}$ that projects the DFT algebroid to the Courant algebroid:
\begin{align*}
{\rm DFT:}& & {\sf L}_{-1}&=C^\infty(\cM)& &\oplus & {\sf L}_{0}&=\Gamma(L)&&\oplus&&{\sf L}_{2}\\
\phi\downarrow\quad&&\phi_1&\,\downarrow && & \phi_1&\,\downarrow &&&\phi_1\,\,&\!\downarrow\\
{\rm CA:}&& {\sf L}'_{-1}&=C^\infty(M)& &\oplus & {\sf L}'_{0}&=\Gamma(E)&&\oplus&&\varnothing~.
\end{align*}
This basically means that if we pick a coordinate chart on $\cal M$, $x^A=(x^a,\tilde{x}_a)$, such that the coordinates $x^a$ correspond to coordinates of the manifold $M$ and $M$ is then implicitly defined by $\tilde{x}_a=\text{const.}$, all functions $f(x^A)$ on $\cal M$ upon restriction only depend on half the coordinates, namely $f(x^a,\tilde{x}_a=\text{const.})$. However, the fiber structure remains unchanged. To verify that such a mapping is indeed an $L_\infty$-morphism one must check that  it satisfies the conditions od Def. \ref{linfmorph}. We begin with $n=0$ that implies only $\phi_1(\mu_0)=0$ as a Courant algebroid does not include spaces ${\sf L}'_1$ nor ${\sf L}'_2$. Therefore $\phi_0=0$ and we are dealing with flat $L_\infty$-morphism. For the case of $n=1$ we   make the following choice:
\begin{align}\label{eq:phi1}
\phi_1(f)&=\sfrac 1 2 f\big|_M~,\\
\phi_1(e)&=e\big|_M~,
\end{align}
where $e\big|_M$ means the component function is restricted and the section exists only over $M$. This case has only one non-trivial identity, the one corresponding to $l=f$:
\[\phi_1(\mu_1(f))-\phi_2(\mu_0,f)=\mu_1'(\phi_1(f))~.\]
By plugging in the products from \ref{subsec:noL1} and Example 3.3, this becomes:
\[\phi_1(\cD f)-\phi_2(\mu_0,f)=D\phi_1(f)~.\]
In $(x,\tilde{x})$ coordinates the DFT anchor splits into two: $\rho^A{}_I=(\rho^a{}_I,\tilde{\rho}_{aI})$, the first is the one we relate to the anchor $a^a{}_I$ of a Courant algebroid. This choice is consistent since for a DFT algebroid we have:
 \[\rho^a{}_I\hat\eta^{IJ}\rho^b{}_J=0~,\label{eq:rhorho2}\]
 according to property 1 of Def. \ref{dftalg1} (see also \eqref{eq:rhorho}), meaning $\rho^a{}_I$ satisfies the first identity in \eqref{Cbox}. Therefore, the DFT derivative splits into two: $\cD=\sfrac 1 2 D+\sfrac 1 2  \tilde{D}$, the first of which we associate with the Courant algebroid differential as its image is in the kernel of $a$. One may wonder about this extra factor of 1/2, it stems from the different definitions of the derivative and pairing in DFT and in a Courant algebroid. Whereas in DFT the derivative carries the factor, in a Courant algebroid the pairing does instead. By using \eqref{eq:phi1} and the fact that $(\partial_a f)\big|_M=\partial_a (f\big|_M)$ and $(\tilde{\partial}^a f)\big|_M\neq\tilde{\partial}^a(f\big|_M)$, one obtains:
\begin{equation}
\phi_2(\mu_0,f)=\sfrac 1 2 (\tilde{D}f)\big|_M~.
\end{equation}

The case of $n=2$  has two non-trivial possibilities: $(l_1,l_2)=\{(f,e),(e_1,e_2)\}$. To keep calculations as simple as possible we shall make the Ansatz that all components $\phi_i$ for $i>1$ not including $\mu_0$ vanish. The first produces:
\[
\phi_3(\mu_0,f,e)-\phi_1(\langle e,\cD f\rangle)=-\langle\phi_1(e),D\phi_1(f)\rangle_C~,
\]
that reduces to the definition:
\begin{equation}\label{eq:phi3mu0fe}
\phi_3(\mu_0,f,e)=\sfrac 1 4 \langle e, \tilde{D}f\rangle\big|_M~.
\end{equation}
The second identity is (again, after the choice of $\phi_2(e_1,e_2)=0$):
\[\phi_3(\mu_0,e_1,e_2)+\phi_1(\cbral e_1,e_2\cbrar)=[\phi_1(e_1),\phi_1(e_2)]_C~.\]
Here we shall make the identification of $\hat{T}$ of DFT with the twist  of the Courant algebroid, therefore trivially one has:
\begin{equation}\label{eq:phi3mu0ee}
\phi_3(\mu_0,e_1,e_2)=[e_1,e_2]_C\big|_M-\cbral e_1,e_2\cbrar\big|_M~.
\end{equation}

In the case of $n=3$ we have three possibilities: $(l_1,l_2,l_3)=\{(e_1,e_2,e_3),(\mu_0,f,e),$ $(\mu_0,f_1,f_2)\}$, the first being a definition as  before and the other two being  consistency checks. The first produces the definition of $\phi_4(\mu_0,e_1,e_2,e_3)$:
\begin{equation}
\phi_4(\mu_0,e_1,e_2,e_3)=\big(\sfrac 1 2 \cN(e_1,e_2,e_3)-\cN_c(e_1,e_2,e_3)\big)\big|_M~,
\end{equation}
The next case is the following consistency condition:
\begin{align*}
-\phi_3(\cD f,\mu_0,e)+\phi_2(\langle e,\cD f\rangle,\mu_0)-\phi_1(\cbral e,\cD f\cbrar-\cD\langle e,\cD f\rangle)=\\
=D\phi_3(\mu_0,f,e)-[\phi_1(e),\phi_2(\mu_0,f)]_C~,
\end{align*}
which, by use of \eqref{eq:phi3mu0fe} and \eqref{eq:phi3mu0ee}, transforms into:
\[ D\langle e,D f\rangle_C-[e,D f]_C=0~.\]
This is valid by Prop. 4.2 of \cite{rw} as all present structures  correspond to a Courant algebroid. The third non-vanishing condition gives:
\begin{align*}
\phi_3(\mu_0,\cD  f_1,f_2)+\phi_3(\mu_0,\cD  f_2,f_1)+\phi_1(2\langle \cD f_1,\cD f_2\rangle)=\\
=\langle\phi_2(\mu_0,f_2),D\phi_1(f_1)\rangle_C+\langle\phi_2(\mu_0,f_1),D\phi_1(f_2)\rangle_C~,
\end{align*}
that vanishes by virtue of $\langle D f_1,D f_2\rangle_C=0$. 

For $n=4$ there are two non-trivial possibilities for the selection of elements: $(l_1,l_2,l_3,l_4)=$ $\{(\mu_0,e_1,e_2,e_3),(\mu_0,e_1,e_2,f)\}$, both producing compatibility conditions. The former combination yields condition:
\begin{align*}
\phi_3(\cbral e_1,e_2\cbrar,\mu_0,e_3)+\text{cyclic(1,2,3)}+\phi_2(\cN(e_1,e_2,e_3),\mu_0)-\phi_1(\mathsf{SC}_{\text{Jac}}(e_1,e_2,e_3))=\\
=D\phi_4(\mu_0,e_1,e_2,e_3)-([\phi_1(e_1),\phi_3(\mu_0,e_2,e_3)]_C+\text{cyclic(1,2,3)})~,
\end{align*}
that is satisfied by use of \eqref{preJ} and the third relation in \eqref{Cbox}.
The condition corresponding to the latter selection of elements is:
\begin{align*}
&\phi_3(\langle e_2,\cD f\rangle,\mu_0,e_1)-e_1\leftrightarrow e_2+\phi_3(\cbral e_1,e_2\cbrar,\mu_0,f)+\phi_4(\cD f,\mu_0,e_1,e_2)=\\
&=\langle\phi_1(e_2),D\phi_3(\mu_0,e_1,f)\rangle_C-e_1\leftrightarrow e_2+{}\\
&\hphantom{\,=\,}+\langle \phi_3(\mu_0,e_1,e_2),D \phi_1(f)\rangle_C+\cN_c (\phi_1(e_1),\phi_1(e_2),\phi_2(\mu_0,f))~,
\end{align*}
reducing to:
\[2\cN_c(D f,e_1,e_2)\big|_M=\langle[e_1,e_2]_C,D f\rangle_C\big|_M~,\]
satisfied by Lemma 5.1 of \cite{rw}. 

Finally, $n=5$ has only one condition to consider, $(l_1,l_2,l_3,l_4,l_5)=(\mu_0,e_1,e_2,e_3,e_4)$:
\begin{align*}
&\phi_4(\mu_0,\cbral e_1,e_2\cbrar,e_3,e_4)+\phi_3(\mu_0,\cN(e_1,e_2,e_3),e_4)+\phi_1(\langle \mathsf{SC}_{\text{Jac}}(e_1,e_2,e_3),e_4\rangle_+)+{}\\
&+\text{antisymm.(1,2,3,4)}={}\\
&=\langle\phi_1(e_1),D\phi_4(\mu_0,e_2,e_3,e_4)\rangle_C+\cN_c (\phi_1(e_1),\phi_1(e_2),\phi_3(\mu_0,e_3,e_4))+\text{antisymm.(1,2,3,4)}~,
\end{align*}
satisfied by the last identity of Lemma \ref{eq:RoyProp} and Lemma 5.2 of \cite{rw}. To summarise we present all non-vanishing morphism components in the following boxed set of equations.
\begin{empheq}[box=\ovalbox]{align*}
\phi_1(f)&=\sfrac 1 2 f\big|_M\\
\phi_1(e)&=e\big|_M\\
\phi_2(\mu_0,f)&=\sfrac 1 2 (\tilde{D}f)\big|_M\\
\phi_3(\mu_0,f,e)&=\sfrac 1 4 \langle e, \tilde{D}f\rangle\big|_M\\
\phi_3(\mu_0,e_1,e_2)&=[e_1,e_2]_C\big|_M-\cbral e_1,e_2\cbrar\big|_M\\
~\phi_4(\mu_0,e_1,e_2,e_3)&=\big(\sfrac 1 2 \cN(e_1,e_2,e_3)-\cN_c(e_1,e_2,e_3)\big)\big|_M~~
\end{empheq}

We finish this section by presenting a minimal extension of the morphism above in order to encompass the algebras of subsection \ref{subsec:withl1} and the extended Courant algebroid $L_\infty$-algebra of \cite{p2}. To accomplish this we must make certain assumptions about this morphism. Our choice is the following:
\begin{itemize}
\item $\phi_1$ morphism components are:
\begin{align*}
\phi_1:\begin{cases}
C^\infty({\cal M})\to C^\infty(M), & f\mapsto\sfrac 1 2 f\big|_M\\
\Gamma(L)\to\Gamma(E), & e\mapsto e\big|_M\\
\mathfrak{X}(\cM)\to\mathfrak{X}(M), & h^A\partial_A\mapsto h^a\big|_M\partial_a
\end{cases}~,
\end{align*}
\item all morphism components constructed above remain unchanged,
\item the only non-vanishing $\phi_i$ without $\mu_0$ as an argument are $\phi_1$,
\item the morphism is ``flat'' i.e. $\phi_0=0$.
\end{itemize}
Details of the calculation of the morphism conditions can be found in appendix \ref{app:calcmorph}. Here we simply state the maps that constitute an $L_\infty$-morphism from a DFT algebroid to a Courant algebroid (both viewed as $L_\infty$-algebras) in the following boxed set of expressions.
\begin{empheq}[box=\ovalbox]{align*}
\phi_1(f)&=\sfrac 1 2 f\big|_M\\
\phi_1(e)&=e\big|_M\\
\phi_2(\mu_0,f)&=\sfrac 1 2 (\tilde{D}f)\big|_M\\
\phi_3(\mu_0,f,e)&=\sfrac 1 4 \langle e, \tilde{D}f\rangle\big|_M\\
\phi_3(\mu_0,e_1,e_2)&=[e_1,e_2]_C\big|_M-\cbral e_1,e_2\cbrar\big|_M\\
\phi_4(\mu_0,e_1,e_2,e_3)&=\big(\sfrac 1 2 \cN(e_1,e_2,e_3)-\cN_c(e_1,e_2,e_3)\big)\big|_M\\\cdashline{1-2}
\phi_1(h)&=h^a\big|_M\partial_a\\
~\phi_{i+2}(h_1,\ldots,h_i,\mu_0,e)&=\big(h_1^{A_1}\cdots h_i^{A_i}\partial_{A_1}\cdots\partial_{A_i}\rho(e)^b-h_1^{a_1}\cdots h_i^{a_i}\partial_{a_1}\cdots\partial_{a_i}a(e)^b\big)\big|_M\partial_b~
\end{empheq}

\section{Concluding remarks}

A DFT algebroid is a geometric structure describing properties of  the C-bracket  relevant for the gauge symmetry of double field theory. Here we discussed its global properties and gave a formulation in terms of an $L_\infty$-algebra.  As indicated in \cite{p3}, this is relevant because  an 
$L_\infty$-algebra can be extended to full, classical  on-shell field theory content  including dynamical fields, equations of motion, Noether identities  (see e.g.\cite{olaf}), by adding corresponding vector spaces. This is, of course, not surprising having in mind that an $L_\infty$-algebra  is actually the geometric structure underlying the BV-BRST complex.  Furthermore, it also can accommodate the off-shell formulation of the theory given by the appropriate  action functional, provided one can define a compatible, non-degenerate,   graded symmetric cyclic inner product needed for defining the variational principle. 

In \cite{p1}, a sigma-model based on the structure of a DFT algebroid was proposed, which, however, was shown to be gauge invariant only up to the strong constraint. Although this result is in accord with standard double field theory,  we would like to go one step further. Our motivation for finding a gauge invariant sigma-model without constraints is based on recent proposals suggesting that  there exist physically relevant closed strings backgrounds which cannot be obtained as a solution of the strong constraint \cite{nc,p1}. Using the $L_\infty$-algebra framework one could bootstrap consistent gauge theories  by choosing the initial data of the theory in the form of 1- and 2-brackets and construct the appropriate higher brackets using the homotopy relations \cite{rb}. This is akin to the deformation of a free gauge theory into an interacting one in the BV-BRST approach, see e.g.\cite{glenn}. 

In the language of graded geometry, an $L_\infty$-algebra is fully defined by a Q-structure, an odd homological vector field on a graded manifold. Here, the graded vector space of a Q-manifold fibered over a point  corresponds precisely to the Chevalley-Eilenberg algebra of an $L_\infty$-algebra. The inner product of \emph{cyclic} $L_\infty$-algebras becomes a compatible symplectic structure on this Q-manifold,  resulting in  a QP-manifold  describing the corresponding  (classical)  field theory.
In this paper we gave the definition of a DFT algebroid in terms of a \textit{curved} $L_\infty$-algebra, implying  that one should be able to formulate a DFT algebroid in terms of a Q-structure, going beyond the results of \cite{sd}.  Still, the existence of a compatible symplectic structure, or cyclic inner product is an open problem we shall address in future work.  A promising route to tackle this issue consists of allowing for a  {\it degenerate} symplectic structure following the construction in \cite{maxim}. In  that case one obtains a presymplectic  generalisation of the BV formalism, which reduces to the standard one after factorising out  the zero modes of the presymplectic  form.  However, for most applications one can employ the presymplectic structure without ever performing  factorization explicitly.  Thus, one  expects to obtain an unconstrained  gauge invariant theory relevant for understanding the implications of T-duality in a field theory setting.

\paragraph{Acknowledgments.} We thank Athanasios Chatzistavrakidis, Maxim Grigoriev,  Olaf Hohm, Branislav Jur\v co and
Christian S\"amann for helpful discussions. Part of the work has been done at the Erwin Schr\"odinger Institute (ESI) in Vienna, during the thematic programme
Higher Structures and Field Theory (Aug-Sep 2020). The work is  supported by the Croatian Science Foundation  project IP-2019-04-4168.
\paragraph{Data availability.}
The data that support the findings of this study are available within the article.

\appendix

\section{Relation of the DFT algebroid with the flux formulation of DFT}\label{app:flux}

Here we review the correspondence between the structural data of a DFT algebroid and double field theory \cite{p1,pp}, using a local basis. Starting from Def. \ref{dftalg1}, we relate the $2d$-dimensional base manifold ${\cal M}$ spanned by $\{X^A\}$, $A=1,\ldots, 2d$ with the doubled configuration space of \DFT\ spanned by $\{x^a, \tilde x_a\}$, $a=1,\ldots,d$. Next, we introduce a local basis for the sections of $L$, $e_I$ where $I=1,\dots,2d$,
with the following operations
\begin{nalign}
		\cbral e_I,e_J\cbrar&=\hat\eta^{KM} \hat T_{IJK}\,e_M~, &\quad
		\langle e_I,e_J\rangle&=\hat\eta_{IJ}~,\\
		\rho(e_I)f&=\rho^A{}_{I}\,\partial_A f~, &\quad
		{\cal D}f={\cal D}^I f\,e_I&=\sfrac 12 \,\rho^A{}_J\,\partial_Af\,\hat\eta^{JI}\,e_{I}
		~,
\end{nalign}
where the generalised 3-form ${\hat T}$ is introduced as a twist of the bracket and  $\hat\eta$ corresponds to the symmetric bilinear form on $L$, with components 
\be\hat{\eta}^{IJ}=\begin{pmatrix}
	0 & \delta^i_j\\
	\delta^j_i & 0
\end{pmatrix},\qquad i,j=1,\ldots,d~.\ee
Property 1 of Def. \ref{dftalg1} implies: 
\be\hat{\eta}^{{I}{J}}\rho^A{}_{{I}}\rho^B{}_{{J}}={\eta^{AB}}~.\label{eq:rhorho} \ee
Using the Leibniz rule, i.e., property 2 of Def. \ref{dftalg1}, one can show that the bracket on a general section $E^I(X)e_I\in \G(L)$ is the C-bracket of \DFT:
\begin{align} \label{cbragen}
\cbral E_1,E_2\cbrar^J= \rho^A{}_I\,\big(E_1^I\,\partial_A
E_2^J-\sfrac 12\, \hat{\eta}^{IJ}\,E_1^K\,\partial_AE_{2K} - E_1\leftrightarrow
E_2\big)+\hat{\eta}^{JM}\hat T_{MIK}\,E_1^I\,E_2^K~,
\end{align}
while  property 3 evaluated in a local basis imposes the antisymmetry of ${\hat T}$ in all three indices.

Relation \eqref{eq:rhorho} enables us to identify the components of the anchor map $\rho$ with the generalised bein ${\cal E}^A{}_{{I}}$ of the flux formulation of DFT, see e.g.\cite{diego}:
\begin{align*}
\hat{\eta}^{{I}{J}}{\cal E}^A{}_{{I}}{\cal E}^B{}_{{J}}={\eta^{AB}}~. 
\end{align*}
 Moreover, properties of the bracket \eqref{prehomo} and \eqref{preJ}  written in a local basis produce:
\begin{align*}
2\rho^{B}{}_{[I}\partial_{\underline{B}}\rho^{A}{}_{J]}-\rho^{A}{}_{M}\hat{\eta}^{MN}\hat T_{NIJ}&={\eta_{BC}\rho^C{}_{[I}\partial^A\rho^B{}_{J]}}~,\\
3\hat{\eta}^{MN} \hat T_{M[JK}\hat T_{IL]N}+4\rho^{A}{}_{[L}\partial_{\underline A}  \hat T_{JKI]}&={{\cal Z}_{JKIL}}~,
\end{align*}
where:
\begin{align*}
\mathcal{Z}_{IJKL}=3\eta_{AD}\eta_{BE}\eta^{CF}\rho^D{}_{[I}\partial_{\underline{F}}\rho^A{}_J\,\rho^E{}_{K}\partial_{\underline{C}}\rho^B{}_{L]}~.
\end{align*}
By direct comparison with the expression for fluxes and their Bianchi identities  in \DFT,  given as (Ref.\cite{diego},  Eqs.(1.2) and (1.5)):
\begin{align*}
&{\cal F}_{IJK}=3{\cal E}_{[I}{}^A\partial_{\underline A}{\cal E}_J{}^B{\cal E}_{K]B}~,\\
&3\hat\eta^{MN} {\cal F}_{M[JK}{\cal F}_{IL]N}+4 {\cal E}_{[L}{}^{M}\partial_{\underline M}  {\cal F}_{JKI]}= 4{\hat{\cal Z}_{JKIL}}~,
\end{align*}
we observe that the twist of the bracket ${\hat T}$  can be identified with the 3-form flux ${\cal F}$ of \DFT\  and ${{\cal Z}_{JKIL}}=4{\hat{\cal Z}_{JKIL}}$.    The origin of the  totally antisymmetric tensor $\mathcal{Z}_{IJKL}$ has been explained  in \cite{p1}, where  we have shown that at the level of the corresponding 3d DFT sigma-model one can realise this term as a Wess-Zumino term on an extension of the membrane worldvolume to four dimensions, as in \cite{th}. However, this distinction is not crucial  in the present context, and in the rest of this paper $\mathcal{Z}_{IJKL}$ is packaged into $\mathsf{SC}_{\rm Jac}$ together with the rest of the strong-constraint breaking terms appearing in the expression for the Jacobiator of the C-bracket \eqref{preJ}.

\section{Homotopy conditions for a DFT algebroid $L_\infty$-algebra when $n\geqslant4$}\label{app:l1}

Taking into account all of the argumentation in section \ref{subsec:withl1}, for arbitrary $n$ one can have only 4 non-trivial distinctly new types of homotopy relations, those that are in space $\mathsf{L}_1$. They will be higher orders of expressions found in the $n=4$ case: $(l_1,\ldots,l_n)=\{(h_1,\ldots,h_{n-1},f),(h_1,\ldots,h_{n-2},e_1,e_2),(\mu_0,h_1,\ldots,h_{n-3},e,f),(\mu_0,h_1,\ldots,h_{n-4},e_1,e_2,e_3)\}$, the first two being the definitions of \eqref{eq:mu0fmu0ee} and the last two consistency conditions. In order we have:
\[\mu_n(h_1,\ldots,h_{n-1},\cD f)=(-1)^{n+1}\mu_{n+1}(\mu_0,h_1,\ldots,h_{n-1},f)~,\]
from which one immediately sees the first line of \eqref{eq:mu0fmu0ee}, and:
\begin{align*}
(-1)^n\mu_{n-1}(\mu_2(e_1,e_2),h_1,\ldots,h_{n-2})+{}\\
+\cdots-{}\\
-(\mu_{n-m}(\mu_{m+1}(h_1,\ldots,h_m,e_1),h_{m+1},\ldots,h_{n-2},e_2)-e_1\leftrightarrow e_2)+{}\\
+\cdots={}\\
=(-1)^{n+1}\mu_{n+1}(\mu_0,h_1,\ldots,h_{n_2},e_1,e_2)~,
\end{align*}
where the dots indicate summation over $m$ and terms of all unshuffles $\sigma$: $h_{\sigma(1)},\ldots,h_{\sigma(m)}$ and  $h_{\sigma(m+1)},\ldots,h_{\sigma(n-2)}$. This summation is nothing more than the product rule expansion of differential operator $h^{A_1}_1\cdots h^{A_{n-2}}_{n-2}\partial_{A_1}\cdots\partial_{A_{n-2}}$ acting on $[\rho(e_1),\rho(e_2)]$ implying the second line of \eqref{eq:mu0fmu0ee}. Continuing on now to the conditions, the third combination of elements produces:
\begin{align*}
(-1)^{n-1}\mu_{n}(\cD f,\mu_0,h_1,\ldots,h_{n-3},e)-\mu_{n-1}(\langle e,\cD f\rangle, \mu_0, h_1,\ldots,h_{n-3})&+{}\\
+\mu_{n-2}(\mu_3(\mu_0,e,f),h_1,\ldots,h_{n-3})&+{}\\
+\cdots&+{}\\
+(-1)^{m+1}\mu_{n-m}(\mu_{n+1}(h_1,\ldots,h_m,e),\mu_0,h_{m+1},\ldots,h_{n-3},f)+\cdots&+{}\\
+(-1)^{n-m}\mu_{n-m}(\mu_{n+1}(\mu_0,h_1,\ldots,h_{m-1},f),h_m,\ldots,h_{m-3},e)+\cdots&+{}\\
+\cdots&=0~,
\end{align*}
where, just as previously, dots indicate summation over $m$ and term of all unshuffles of $h$. Again, realising this summation can be resummed produces:
\[(-1)^nh^{A_1}_1\cdots h^{A_{n-3}}_{n-3}\partial_{A_1}\cdots\partial_{A_{n-3}}(-\mathsf{SC}_\rho(e,\cD f)^D-\sfrac 1 4 \eta^{DC}\partial_C(\rho(e)f)+\sfrac 1 2 \eta_{CD}\rho(e)^B\partial_B\partial_Cf)\partial_D=0~,\]
that vanishes in the exact same way it does in the $n=3,4$ cases. The final condition valid for $n\geqslant 4$ is:
\begin{align*}
(-1)^n\mu_{n-1}(\mu_2(e_1,e_2),\mu_0,h_1,\ldots,h_{n-4},e_3)+\text{cyclic}&+{}\\
+\mu_{n-2}(\mu_3(e_1,e_2,e_3),\mu_0,h_1,\ldots,h_{n-4})&+{}\\
+\mu_{n-3}(\mu_4(\mu_0,e_1,e_2,e_3),h_1,\ldots,h_{n-4})&+{}\\
+\cdots&+{}\\
+(-1)^{n-m-1}\mu_{n-m}(\mu_{m+1}(\mu_0,h_1,\ldots,h_{m-2},e_1,e_2),h_{m-1},\ldots,h_{n-4},e_3)+\text{cyclic}+\cdots&+{}\\
+(-1)^m\mu_{n-m}(\mu_{m+1}(h_1,\ldots,h_m,e_1),\mu_0,h_{m+1},\ldots, h_{n-4},e_2,e_3)+\text{cyclic}+\cdots&+{}\\
+\cdots&=0~,
\end{align*}
utilising the same logic as before it is easy to see this is nothing more than a higher derivative of the condition obtained for $n=4$:
\begin{align*}(-1)^nh^{A_1}_1\cdots h^{A_{n-4}}_{n-4}\partial_{A_1}\cdots\partial_{A_{n-4}}\big(-\mathsf{SC}_\rho(\cbral e_1,e_2\cbrar,e_3)-[\mathsf{SC}_\rho(e_1,e_2),\rho(e_3)]+\text{cycl.}&-{}\\
-\rho\,\text{Jac}(e_1,e_2,e_3)\big)&=0~,\end{align*}
again satisfied by the fact that $\text{Jac}(\rho(e_1),\rho(e_2),\rho(e_3))=0$.

\section{Calculation of morphism conditions with degree 1 spaces}\label{app:calcmorph}

In this appendix we show the calculation of morphism conditions in the case of DFT and Courant algebroids with spaces ${\sf L}_1$ (respectively ${\sf L}_1'$). In order to calculate the remaining morphism components we turn to the condition of Def. \ref{linfmorph} order by order, beginning with $n=0$ that does not change with respect to section \ref{subsec:morphnol1} by the introduction of space ${\sf L}_1$. Next, $n=1$, has only one new and non-trivial condition:
\[
\phi_1(\rho(e))-\phi_2(\mu_0,e)=a(\phi_1(e))~,
\]
implying:
\begin{equation}\label{eq:phi2mu0e}
\phi_2(\mu_0,e)=0~.
\end{equation}
In the case of $n=2$, by taking into account the assumptions above we have three new and non-trivial morphism conditions. The first is a compatibility condition corresponding to $(l_1,l_2)=(\mu_0,f)$:
\[
\phi_1\big(\sfrac 1 2  \eta^{-1}(df)\big)=a\big(\sfrac 1 2 \tilde{D}f\big|_M\big)~,
\]
that is satisfied by \eqref{eq:rhorho2} and \eqref{eq:eta}. Next, for $(l_1,l_2)=(h,e)$, the morphism condition produces:
\[
\phi_3(\mu_0,h,e)+\phi_1(h^A\partial_A\rho(e)^B\partial_B)=\phi_1(h)^a\partial_aa(\phi_1(e))^b\partial_b~,
\]
by splitting capital indices one obtains:
\[
\phi_3(\mu_0,h,e)=-(\tilde{h}_a\tilde{\partial}^a\rho(e)^b)\big|_M\partial_b~.
\]
Lastly for $n=2$ we have $(l_1,l_2)=(h,f)$ that results in:
\[\phi_3(\mu_0,h,f)=0~,\]
directly from our assumptions and Def. \ref{linfmorph}.
Before moving on to higher cases of $n$, guided by what we have obtained thus far for $n=1,2$ we shall make the following Ansatz for the new components to $\phi$:
\be\label{eq:phimu0e}
\phi_{i+2}(h_1,\ldots,h_i,\mu_0,e)=\big(h_1^{A_1}\cdots h_i^{A_i}\partial_{A_1}\cdots\partial_{A_i}\rho(e)^b-h_1^{a_1}\cdots h_i^{a_i}\partial_{a_1}\cdots\partial_{a_i}a(e)^b\big)\big|_M\partial_b~,
\ee
or explicitly:
\begin{align*}
\phi_{i+2}(h_1,\ldots,h_i,\mu_0,e)&=\big( \tilde{h}_{1a_1}\cdots\tilde{h}_{ia_i}\tilde{\partial}^{a_1}\cdots\tilde{\partial}^{a_i}\rho(e)^b+h_1^{a_1}\tilde{h}_{2a_2}\cdots\tilde{h}_{ia_i}\partial_{a_1}\tilde{\partial}^{a_2}\cdots\tilde{\partial}^{a_i}\rho(e)^b+\\
&\phantom{\,=\,\big( }+\cdots\big)\big|_M\partial_b~,
\end{align*}
where the dots indicate all possible combinations of $h$ and $\tilde{h}$ except the one with no $\tilde{h}$. All other possible new $\phi$ components vanish. Continuing on to $n=3$ where after taking into account our assumptions and previous Ansatz one finds there are, in fact, three new and non-trivial identities to be satisfied. The first, $(l_1,l_2,l_3)=(h_1,h_2,e)$, is just the definition \eqref{eq:phimu0e}, however the second $(l_1,l_2,l_3)=(\mu_0,h,f)$ and third $(l_1,l_2,l_3)=(\mu_0,e_1,e_2)$ are consistency conditions of our Ansatz and yield, respectively:
\begin{align}\label{eq:n3ident}
\begin{split}
-\phi_3(\cD f,\mu_0,h)-\phi_1(\sfrac 1 2 \eta^{BC}h^A\partial_A\partial_Bf\partial_C)=- \phi_1(h)^a\partial_a(\phi_2(\mu_0,f))^b\partial_b~,\\
-\phi_3(\rho(e_1),\mu_0,e_2)+\phi_3(\rho(e_2),\mu_0,e_1)+\phi_1({\sf SC}_\rho(e_1,e_2))=a(\phi_3(\mu_0,e_1,e_2))~.
\end{split}
\end{align}
The latter is satisfied automatically once one takes into account $\rho\circ\tilde{D}=\tilde{\partial}$, whereas for the former one needs the homomorphism property of the Courant bracket and relations \eqref{prehomo}. It may be useful to note that since \eqref{eq:phimu0e} for $i=0$ vanishes as is in accord with \eqref{eq:phi2mu0e} the second relation above does not have an extra term coming from \eqref{eq:phimu0e} that will appear in higher identities. 

The analysis so far enables us to move on to the case of a general $n$. As $\phi$ only has infinite components for one combination of elements, $\phi_{i+2}(h_1,\ldots,h_i,\mu_0,e)$, we need only look at identities involving this component since all others are taken care of in explicit $n$ cases either above or in section \ref{subsec:morphnol1} for identities that are equivalent. That withstanding we have three possibilities, starting with $(l_1,\ldots,l_n)=(h_1,\ldots,h_{n-1},e)$ which is simply the definition \eqref{eq:phimu0e}. Next, $(l_1,\ldots,l_n)=(\mu_0,h_1,\ldots,h_{n-2},f)$, is simply the higher derivative case of the first line in \eqref{eq:n3ident} vanishing for the same reason. Finally, the generalisation of the second line or $(l_1,\ldots,l_n)=(\mu_0,h_1,\ldots,h_{n-3},e_1,e_2)$ and the only slightly non-trivial identity for a general $n$. Definition \ref{linfmorph} implies:
\begin{align*}
&\cdots+{}\\
&+(-1)^j\phi_{n-j+1}(\mu_j(h_1,\ldots,h_{j-1},e_1),\mu_0,h_j,\ldots,h_{n-3},e_2)-{}\\
&-(-1)^j\phi_{n-j+1}(\mu_j(h_1,\ldots,h_{j-1},e_2),\mu_0,h_j,\ldots,h_{n-3},e_1)+{}\\
&+\cdots+{}\\
&+(-1)^n\phi_{n-1}(\mu_2(e_1,e_2),\mu_0,h_1,\ldots,h_{n-3})+\phi_1(\mu_n(\mu_0,h_1,\ldots,h_{n-3},e_1,e_2))={}\\
&=\cdots+{}\\
&\phantom{\,=\,}+(-1)^{n-j-1}\mu'_{n-j+1}(\phi_j(\mu_0,h_1,\ldots,h_{j-2},e_1),\phi_1(h_1),\ldots,\phi_1(h_{n-3}),\phi_1(e_2))-{}\\
&\phantom{\,=\,}-(-1)^{n-j-1}\mu'_{n-j+1}(\phi_j(\mu_0,h_1,\ldots,h_{j-2},e_2),\phi_1(h_1),\ldots,\phi_1(h_{n-3}),\phi_1(e_1))+{}\\
&\phantom{\,=\,}+\cdots+{}\\
&\phantom{\,=\,}+\mu'_{n-2}(\phi_3(\mu_0,e_1,e_2),\phi_1(h_1),\ldots,\phi_1(h_{n-3}))~,
\end{align*}
where by resumming the partial derivatives and utilising the definitions of maps $\phi_i$, $\mu_i$, and $\mu'_i$ one obtains
\begin{align*}
&h_1^{A_1}\cdots h_{n-3}^{A_{n-3}}\partial_{A_1}\cdots\partial_{A_{n-3}}([\rho(e_1),\rho(e_2)]^b-{\sf SC}_\rho(e_1,e_2)^b-\rho\cbral e_1,e_2\cbrar^b)\partial_b={}\\
&=h_1^{a_1}\cdots h_{n-3}^{a_{n-3}}\partial_{a_1}\cdots\partial_{a_{n-3}}([a(e_1),a(e_2)]^b-a[e_1,e_2]_C^b)\partial_b~,
\end{align*}
with ${\sf SC}_\rho(e_1,e_2)^b$ defined by the splitting $\partial_B=\partial_b+\tilde{\partial}^b$. Each side of this equality vanishes on its own, the lhs because of \eqref{prehomo} and the rhs because of the homomorphism property of a Courant algebroid anchor map $a$.

\end{document}